\documentclass[12pt]{elsart}
\usepackage{graphicx}
\usepackage{amssymb}
\usepackage{natbib}
\usepackage{amsmath}

\begin{document}

\begin{frontmatter}
\title{Monogamy and entanglement in tripartite quantum states}
\author{Chang-shui Yu,\corauthref{yu}}
\corauth[yu]{\emph{Email address:} quaninformation@sina.com;
\emph{or} ycs@dlut.edu.cn}
\author{ He-shan Song}
\address{School of Physics and Optoelectronic Technology,
Dalian University of Technology, Dalian 116024, P. R. China.
Phone:86-0411-84706201. Fax:86-0411-84709304.}
\date{Received: date / Revised version: date}

\begin{abstract}
 We present an interesting monogamy equation for $\left( 2\otimes
2\otimes n\right) $-dimensional pure states, by which a quantity is
found to characterize the tripartite entanglement with the GHZ type
and W type entanglements as a whole. In particular, we, for the
first time, reveals that for any quantum state of a pair of qubits,
the difference between the two remarkable entanglement measures,
concurrence and negativity, characterizes the W type entanglement of
tripartite pure states with the two-qubit state as reduced density.
\end{abstract}

\end{frontmatter}

\section{Introduction}

Entanglement or the nonseparability of quantum states of composite systems
is an essential feature of quantum mechanics and plays crucial role in
various applications in quantum information processing [1-4]. It is of a
paramount importance and one of the main tasks of quantum entanglement
theory to quantitively characterize the extent to which composite quantum
systems are entangled by constructing a so-called entanglement measure (a
mathematical function) that should not increase on averaging under local
operations and classical communications (LOCC). However, only the
entanglements of bipartite pure states and low-dimensional systems are well
understood [5-7]. Needless to say, it is still a challenge to quantify
entanglement for high-dimensional systems and multipartite mixed states, the
quantification of entanglement is not well solved even for multipartite pure
states [8]. One of the main reasons is that multipartite entanglement can be
classified into many inequivalent classes [9-11]. Therefore, we have to
concern which class the multipartite entanglement belongs to when we
consider multipartite entanglement.

In recent years, several works have shown that multipartite entanglements of
some kinds have close contact with the monogamy of entanglement, a key
property of entanglement which, quite different from classical correlation,
demonstrates that the degree to which either of two parties can be entangled
with anything else seems to be constrained by the entanglement that may
exist between the two quantum parties [12-14]. A remarkable example [15] is
that the residual entanglement of the Coffman-Kundu-Wootters (CKW)
inequality characterizes the GHZ type entanglement of tripartite pure states
of qubits. The same residual entanglement can also be obtained from the
monogamy inequality dual to CKW inequality based on concurrence of
assistance (COA) [16] presented for tripartite systems of qubits by Gour et
al [17]. Quite recently, we have found an interesting monogamy equation for $%
\left( 2\otimes 2\otimes n\right) $-dimensional (or multiple qubits) quantum
pure states which relates the bipartite concurrence, COA and GHZ type
tripartite entanglement [18,19]. In this paper, we first present a new
monogamy equation for $\left( 2\otimes 2\otimes n\right) $-dimensional (or
multiple qubits) quantum pure states. The distinct advantage is that the
"residual quantity" deduced from the equation can characterize the
tripartite entanglement with GHZ type and W type entanglements as a whole.
Furthermore, it is invariant under local unitary transformations and does
not increase under the local operations performed on the $n$- dimensional
subsystem, hence it is an entanglement semi-monotone.

What is more, it is well-known to all that there are two remarkable
entanglement measures for bipartite systems of qubits-------- One is the
concurrence [20] and the other is the negativity [21,22]. It has been shown
that the negativity is not greater than the concurrence [23]. But what does
the difference between the two entanglement measures imply? In this paper,
based on the new monogamy equation we find a surprising fact that the
tripartite W type entanglement of a $\left( 2\otimes 2\otimes n\right) $%
-dimensional quantum pure state can be characterized by the difference
between concurrence and negativity of the ($2\otimes 2)$ -dimensional
reduced density matrix. This paper is organized as follows. We first
introduce our interesting monogamy inequality for tripartite pure states;
Then we reveals that the difference between concurrence and negativity can
characterize the W type entanglement; The conclusion is drawn finally.

\section{Monogamy and entanglement for tripartite pure states}

Let us first briefly recall the concurrence and the negativity for the
bipartite quantum state $\varrho $ of qubits. The concurrence is defined as%
\begin{equation}
C\left( \varrho \right) =\max \{0,\lambda _{1}-\sum\limits_{i>1}\lambda
_{i}\},
\end{equation}%
where $\lambda _{i}$ are the square roots of the eigenvalues of $\varrho
\tilde{\varrho}$ in decreasing order with $\tilde{\varrho}=\left( \sigma
_{y}\otimes \sigma _{y}\right) \varrho ^{\ast }\left( \sigma _{y}\otimes
\sigma _{y}\right) $ and $\sigma _{y}$ is the Pauli matrix, and the
negativity is defined as [22]%
\begin{equation}
N\left( \varrho \right) =\left\Vert \varrho ^{T_{\alpha }}\right\Vert _{1}-1
\end{equation}%
which corresponds to the doubled absolute value of the sum of the negative
eigenvalues of $\varrho ^{T_{\alpha }}$, where $\varrho ^{T_{\alpha }}$
denotes the partial transpose of $\varrho $ and $\left\Vert \cdot
\right\Vert $ is the trace norm of a matrix. It has been shown [23] that
\begin{equation}
C\left( \varrho \right) \geq N\left( \varrho \right)
\end{equation}%
where the $"="$ always holds for pure $\varrho $ and some given mixed states
[24].

Now suppose that a tripartite $\left( 2\otimes 2\otimes n\right) $-
dimensional pure state $\left\vert \Psi \right\rangle _{ABC}$ is shared by
three parties Alice, Bob and Charlie where the subsystem at Charlie's side
is an auxilliary one. The $\left( 2\otimes 2\right) $- dimensional reduced
density matrix by tracing over party C can be given by $\rho _{AB}=Tr_{C}$ $%
\left( \left\vert \Psi \right\rangle _{ABC}\left\langle \Psi \right\vert
\right) $. Based on GHJW theorem [25,26], any decomposition of $\rho _{AB}$
can always be realized with the help of Charlie who can perform
positive-operator-value-measurements [27] on his subsystem C. Let $\mathcal{E%
}=\{p_{i},\left\vert \varphi _{i}^{AB}\right\rangle \}$ is a decomposition
of $\rho _{AB}$ such that
\begin{equation}
\rho _{AB}=\sum\limits_{i}p_{i}\left\vert \varphi _{i}^{AB}\right\rangle
\left\langle \varphi _{i}^{AB}\right\vert ,\sum\limits_{i}p_{i}=1.
\end{equation}%
Oppositely to the mixed-state concurrence which is defined by the minimal
average pure-state concurrence on $\mathcal{E}$, the COA is defined [28,29]
as
\begin{eqnarray}
C_{a}\left( \left\vert \Psi \right\rangle _{ABC}\right) &=&\max_{\mathcal{E}%
}\sum\limits_{i}p_{i}C\left( \left\vert \varphi _{i}^{AB}\right\rangle
\right) \\
=C_{a}\left( \rho _{AB}\right) &=&{tr}\sqrt{\sqrt{\rho _{AB}}\tilde{\rho}%
_{AB}\sqrt{\rho _{AB}}}=\sum\limits_{i=1}^{4}\lambda _{i},
\end{eqnarray}%
where Charlie is to maximize the entanglement shared by Alice and Bob and
the corresponding parameters are defined the same as eq. (1). Besides, it
should be noted that the COA is a tripartite entanglement monotone instead
of a bipartite one [30]. Quite recently we have found an interesting
monogamy equation\footnote{%
The monogamy relation is different from the familiar monogamy first
introduced in Ref. [9]. However, it does not violate the property of
monogamy, which does show the limitation of entanglement shared between
different parties. In this sense (at least a generalized sense), all that
imply the similar limitations of entanglement are considered as monogamy.},%
\begin{equation}
C_{a}^{2}\left( \rho _{AB}\right) -C^{2}(\rho _{AB})=\tau ^{2}(\left\vert
\Psi \right\rangle _{ABC}),
\end{equation}%
where $\tau (\left\vert \Psi \right\rangle _{ABC})$ is a good entanglement
measure for GHZ type entanglement. Since eq. (3) always holds for the states
of two qubits, if one replaces the concurrence by the negativity, eq. (7)
can lead to another monogamy equation as%
\begin{equation}
C_{a}^{2}\left( \rho _{AB}\right) -N^{2}(\rho _{AB})=\chi ^{2}(\left\vert
\Psi \right\rangle _{ABC}).
\end{equation}%
Now we claim

\textbf{Theorem 1: } For a $\left( 2\otimes 2\otimes n\right) $- dimensional
quantum pure state $\left\vert \Psi \right\rangle _{ABC}$, $\chi (\left\vert
\Psi \right\rangle _{ABC})$ is an entanglement semi-monotone that is
invariant under local unitary transformations and does not increase under
the local operations performed on the $n$- dimensional subsystem. $\chi
(\left\vert \Psi \right\rangle _{ABC})$ characterizes tripartite
entanglement with GHZ type and W type entanglement as a whole.

\textbf{Proof.} It is obvious that
\begin{equation}
C_{a}^{2}\left( \rho _{AB}\right) -N^{2}\left( \rho _{AB}\right) \geq 0.
\end{equation}%
Furthermore, according to the definitions of COA and negativity, one can
find that it is impossible to change $\chi (\left\vert \Psi \right\rangle
_{ABC})$ by local unitary transformations on any subsystems. The most
general local operations can be given in terms of Kraus operators denoted by
$M_{k}$, $\sum_{k}M_{k}^{\dag }M_{k}\leq I_{C}$. The state after the
operation $M_{k}$ can be represented by
\begin{equation*}
\left\vert \psi \right\rangle _{k}=\left( I_{AB}\otimes M_{k}\right)
\left\vert \Psi \right\rangle _{ABC}\left\langle \Psi \right\vert \left(
I_{AB}\otimes M_{k}^{\dag }\right) /p_{k},
\end{equation*}%
with $I_{AB}$ being the $\left( 2\otimes 2\right) $-dimensional identity and
$p_{k}=$tr$\left\vert \psi \right\rangle _{kk}\left\langle \psi \right\vert $%
. Thus the average $\bar{\chi}(\cdot )$ can be given by

\begin{eqnarray}
&&\bar{\chi} =\sum_{k}p_{k}\chi (\left\vert \psi \right\rangle _{k})  \notag
\\
&=&\sum_{k}p_{k}\sqrt{C_{a}^{2}\left( \left\vert \psi \right\rangle
_{k}\right) -N^{2}\left( \left\vert \psi \right\rangle _{k}\right) }  \notag
\\
&=&\sum_{k}\sqrt{p_{k}\left[ C_{a}\left( \left\vert \psi \right\rangle
_{k}\right) -N\left( \left\vert \psi \right\rangle _{k}\right) \right] }
\notag \\
&&\times \sqrt{p_{k}\left[ C_{a}\left( \left\vert \psi \right\rangle
_{k}\right) +N\left( \left\vert \psi \right\rangle _{k}\right) \right] }
\notag \\
&\leq &\left\{ \sum_{k}p_{k}\left[ C_{a}\left( \left\vert \psi \right\rangle
_{k}\right) -N\left( \left\vert \psi \right\rangle _{k}\right) \right]
\right\} ^{1/2}  \notag \\
&&\times \left\{ \sum_{k}p_{k}\left[ C_{a}\left( \left\vert \psi
\right\rangle _{k}\right) +N\left( \left\vert \psi \right\rangle _{k}\right) %
\right] \right\} ^{1/2}  \notag \\
&=&\sqrt{\left[ \sum_{k}p_{k}C_{a}\left( \left\vert \psi \right\rangle
_{k}\right) \right] ^{2}-\left[ \sum_{k}p_{k}N\left( \left\vert \psi
\right\rangle _{k}\right) \right] ^{2}}  \notag \\
&\leq &\sqrt{C_{a}^{2}\left( \rho _{AB}\right) -N^{2}\left( \rho
_{AB}\right) }=\chi (\left\vert \Psi \right\rangle _{ABC}).
\end{eqnarray}%
Here the first inequality follows from the Cauchy-Schwarz inequality
\begin{equation}
\left( \sum_{i}x_{i}^{2}\right) ^{1/2}\left( \sum_{j}y_{j}^{2}\right)
^{1/2}\geq \sum_{k}x_{k}y_{k},x,y\geq 0.
\end{equation}%
The second inequality is derived from the definition of COA and the
convexity of negativity. Eq. (10) shows that $\chi (\left\vert \Psi
\right\rangle _{ABC})$ is an entanglement semi-monotone.

Now we show $\chi (\left\vert \Psi \right\rangle _{ABC})$ characterizes the
tripartite entanglement of $\left\vert \Psi \right\rangle _{ABC}$ with the
GHZ type and W type entanglement as a whole. We first show that $\chi
(\left\vert \Psi \right\rangle _{ABC})$ vanishes for separable states and
then show that $\chi (\left\vert \Psi \right\rangle _{ABC})$ has nonzero
value for both GHZ type entanglement and W type entanglement. Without loss
of generality, a separable tripartite pure state can be written as
\begin{equation}
\left\vert \Phi _{1}\right\rangle _{ABC}=\left\vert \psi \right\rangle
_{AB}\otimes \left\vert \phi \right\rangle _{C}
\end{equation}%
or
\begin{equation}
\left\vert \Phi _{2}\right\rangle _{ABC}=\left\vert \phi \right\rangle
_{A}\otimes \left\vert \psi \right\rangle _{BC}
\end{equation}%
where $\left\vert \phi \right\rangle $ represents a pure state of a single
qubit and $\left\vert \psi \right\rangle $ denotes a general bipartite pure
state which may be entangled or not. Thus a fully separable state can be
included in either of eq. (12) and eq. (13). The reduced density matrix $%
\rho _{AB1}$ of $\left\vert \Phi _{1}\right\rangle _{ABC}$ is pure, hence
\begin{equation}
C_{a}\left( \rho _{AB1}\right) =N\left( \rho _{AB1}\right) =C\left( \rho
_{AB1}\right) ,
\end{equation}%
which shows $\chi (\left\vert \Phi _{1}\right\rangle _{ABC})=0$. The reduced
density matrix $\rho _{AB2}$ of $\left\vert \Phi _{2}\right\rangle _{ABC}$
is separable, hence $C_{a}\left( \rho _{AB1}\right) =N\left( \rho
_{AB1}\right) =C\left( \rho _{AB1}\right) =0$, which also shows $\chi
(\left\vert \Phi _{2}\right\rangle _{ABC})=0$. Therefore, $\chi \left( \cdot
\right) $ vanishes for separable states.

Since GHZ type and W type entanglement are distinguished by sLOCC
operations, which shows that quantum states belonging to the same type can
be converted to each other by invertible local operations[9], it is enough
to only consider the standard states corresponding to every type
entanglement. Generically, GHZ type entangled state can be written as [9]%
\begin{equation}
\left\vert \psi _{GHZ}\right\rangle =\lambda _{0}\left\vert 000\right\rangle
+\lambda _{1}e^{i\theta }\left\vert 111\right\rangle ,
\end{equation}%
where $\lambda _{i}>0$, $\sum\limits_{i}\lambda _{i}^{2}=1$, $\theta \in
\lbrack 0,\pi ]$. W type entangled state can be given by [9]%
\begin{equation}
\left\vert \psi _{W}\right\rangle =\tilde{\lambda}_{0}\left\vert
001\right\rangle +\tilde{\lambda}_{1}\left\vert 010\right\rangle +\tilde{%
\lambda}_{2}\left\vert 100\right\rangle +\tilde{\lambda}_{3}\left\vert
000\right\rangle ,
\end{equation}%
where $\tilde{\lambda}_{0},\tilde{\lambda}_{1},\tilde{\lambda}_{2}>0$, $%
\tilde{\lambda}_{3}=\sqrt{1-\sum_{i=0}^{2}\tilde{\lambda}_{i}^{2}}$. It is
easy to see that
\begin{equation}
C_{a}\left( \left\vert \psi _{GHZ}\right\rangle \right) =2\lambda
_{0}\lambda _{1}\neq 0,N\left( \rho _{AB}^{\prime }\right) =0
\end{equation}%
with $\rho _{AB}^{\prime }=$tr$_{C}\left\vert \psi _{GHZ}\right\rangle
\left\langle \psi _{GHZ}\right\vert $ and
\begin{equation}
C_{a}\left( \left\vert \psi _{W}\right\rangle \right) =C\left( \varrho
_{AB}^{\prime }\right) =2\tilde{\lambda}_{1}\tilde{\lambda}_{2},
\end{equation}%
\begin{equation}
N(\varrho _{AB}^{\prime })=\sqrt{\tilde{\lambda}_{0}^{4}+4\tilde{\lambda}%
_{1}^{2}\tilde{\lambda}_{2}^{2}}-\tilde{\lambda}_{0}^{2},
\end{equation}%
with $\varrho _{AB}^{\prime }=$tr$_{C}\left\vert \psi _{W}\right\rangle
\left\langle \psi _{W}\right\vert $. Eq. (17) shows that $\chi \left( \cdot
\right) $ has nonzero value for GHZ state. Eq. (18) and eq. (19) lead to
\begin{equation}
C_{a}\left( \left\vert \psi _{W}\right\rangle \right) =C\left( \varrho
_{AB}^{\prime }\right) >N(\varrho _{AB}^{\prime }),
\end{equation}%
which shows that $\chi \left( \cdot \right) $ has also nonzero value for W
state.

The previous paragraph has shown that $\chi \left( \cdot \right) $ has also
nonzero value for tripartite quantum pure state with local rank (2,2,2)
[11]. The local rank is defined as the rank of the reduced density matrix
traced out for all except one party. It has been shown that $\left( 2\otimes
2\otimes n\right) $- dimensional quantum pure states can be divided into 9
classes in terms of different local ranks. The standard states of the
classes corresponding to high local ranks can be given by%
\begin{eqnarray}
\left\vert \Phi \right\rangle _{223} &=&\left\vert 000\right\rangle
+\left\vert 011\right\rangle +\left\vert 112\right\rangle , \\
\left\vert \Phi ^{\prime }\right\rangle _{223} &=&\left\vert
000\right\rangle +\frac{1}{\sqrt{2}}\left( \left\vert 011\right\rangle
+\left\vert 101\right\rangle +\left\vert 112\right\rangle \right) , \\
\left\vert \Phi \right\rangle _{224} &=&\left\vert 000\right\rangle
+\left\vert 011\right\rangle +\left\vert 102\right\rangle +\left\vert
113\right\rangle ,
\end{eqnarray}%
where we omit the normalized constant and the subscripts denote the local
rank. For example, '223' denotes the local rank is (2,2,3) and so on. A
quantum pure state with high local rank can always be converted into these
standard states with corresponding local rank based on stochastic LOCC
operations. One can also verify that $\chi \left( \cdot \right) $ does not
vanish for the standard states with high local ranks. That is to say, $\chi
\left( \cdot \right) $ vanishes for separable states including partially
entangled and fully separable states and has nonzero value for tripartite
entangled states, hence $\chi \left( \cdot \right) $ characterizes the
tripartite entanglement.

The entanglement monogamy of eq. (7) and eq. (8) is embodied respectively in
limited GHZ type entanglement and the general tripartite entanglement (W
type entanglement is implied) by bipartite entanglement. That is to say,
when the maximal and minimal entanglement shared by two parties are close
enough, the two parties can not entangle with a third party. Since $\chi
\left( \cdot \right) $ in eq. (8) characterizes the tripartite entanglement
and $\tau ^{2}(\cdot )$ in eq. (7) characterizes the GHZ type entanglement,
it is a natural conjecture that $\chi \left( \cdot \right) -\tau ^{2}(\cdot )
$ should characterizes the W type entanglement.

\textbf{Theorem 2.}-For a $\left( 2\otimes 2\otimes n\right) $- dimensional
quantum pure state $\left\vert \Psi \right\rangle _{ABC}$, $\varpi
(\left\vert \Psi \right\rangle _{ABC})$ is defined as%
\begin{equation}
\varpi (\left\vert \Psi \right\rangle _{ABC})=\varpi (\rho _{AB})=C^{2}(\rho
_{AB})-N^{2}(\rho _{AB}),
\end{equation}%
which is invariant under local unitary transformations and characterizes the
W type entanglement, where $\rho _{AB}=Tr_{C}$ $\left( \left\vert \Psi
\right\rangle _{ABC}\left\langle \Psi \right\vert \right) $.

\textbf{\ Proof}. Analogously to Theorem 1, it is obvious that $\varpi
(\left\vert \Psi \right\rangle _{ABC})$ is invariant under local unitary
transformations. One can find that $\varpi (\cdot )$ is zero for separable
states given in eq. (12) and eq. (13). A simple calculation can also show
that the concurrence and the negativity both vanish for the reduced density
matrix of $\left\vert \psi _{GHZ}\right\rangle $. However, for the $%
\left\vert \psi _{W}\right\rangle $ one can find that $\varpi (\rho _{AB})$
is always nonzero in terms of eq. (20). It is interesting that $\varpi
(\cdot ) $ vanishes for the standard states with high local ranks given by
eqs. (21-23). That is to say, $\varpi (\cdot )$ only characterizes the W
type entanglement with local rank $(2,2,2)$, i.e., W type entanglement, even
the standard states with high local ranks can be converted to the W type
entangled states with local rank $(2,2,2)$. In fact, it is not strange. It
has been shown that $\tau (\cdot )$ quantify GHZ type entanglement by
considering GHZ type entanglement with local rank $(2,2,2)$ as minimal unit,
hence the entanglement states with high local ranks has been quantified as
GHZ type entanglement. Thus the contributions of the standard states with
high local ranks has been subtracted from the total tripartite entanglement $%
\chi \left( \cdot \right) $. The remaining is only the W type entanglement
with local rank $(2,2,2)$.

\textbf{Remark.}-For a tripartite $(2\otimes 2\otimes n)$ -dimensional pure
state $\left\vert \Psi \right\rangle _{ABC}$, let
\begin{equation}
\eta =C(\rho _{AB})-N(\rho _{AB}),
\end{equation}
with $\rho_{AB}={tr}_C\left\vert \Psi \right\rangle _{ABC}\left\langle
\Psi\right\vert$, then $\eta $ characterizes W type entanglement.

\textbf{Proof.} The proof is straightforward in terms of theorem 2. One
might wonder why the characterization of W type entanglement including
Theorem 2 is the substraction of two different entanglement measures. In
fact, concurrence of bipartite reduced density can also well distinguish W
type from GHZ type entanglement. But concurrence per se can not distinguish
states with W type entanglement from some separable states.

Before the end, we would like to emphasize that, even though $\chi \left(
\cdot \right) $, $\varpi \left( \cdot \right) $ and $\eta \left( \cdot
\right) $ are not entanglement monotones, but invariant under local unitary
transformations (or only an entanglement semi-monotone), in many cases they
can be safely used because it was shown in Ref. [8] that it is not necessary
for an entanglement measure to be always an entanglement monotone. For
example, when Alice prepares a $\left( 2\otimes 2\otimes n\right) $-
dimensional quantum pure state, and only sends the n-dimensional qudit to
Bob via a quantum channel, Alice and Bob can safely employ $\chi \left(
\cdot \right) $ to study the evolution of the tripartite entanglement. In
addition, we present theorem 2 and the Remark is in order to reveal the
nature of the difference between the concurrence and the negativity instead
of only to present a W type entanglement measure. Of course, $\varpi \left(
\cdot \right) $ and $\eta $ are both invariant under local unitary
transformations, which can also be employed to measure W type entanglement
in some field. Both eq. (8) and eq. (24), as well as eq. (25), can be
generalized to mixed states by extending the involved entanglement measures
of pure states to mixed states in terms of convex roof construction. But
even though the generalized equations can provide the monogamy relationship
of entanglement, the corresponding $\chi \left( \cdot \right) $ or $\varpi
\left( \cdot \right) $ generalized for mixed states can not exactly
characterize tripartite entanglement as for pure states.

\section{Conclusion and discussion}

We have presented an interesting monogamy equation of entanglement by which
a quantity $\chi \left( \cdot \right) $ can be found to characterize
tripartite $\left( 2\otimes 2\otimes n\right) $-dimensional quantum pure
states with the GHZ type and W type entanglements as a whole. In particular,
we find that the W type entanglement of $\left( 2\otimes 2\otimes n\right) $%
-dimensional pure states can be characterized by the difference between the
two remarkable entanglement measures, the concurrence and the negativity of
the $\left( 2\otimes 2\right) $ -dimensional reduced density matrix of the
tripartite pure states. Finally, we have to mention that $\chi \left( \cdot
\right) $ is an entanglement semi-monotone and $\varpi \left( \cdot \right) $
and $\eta $ are both invariant under local unitary transformations [31].
However, they show the interesting relations between different entanglement
measures and reveal some valuable implication after all. It is our
forthcoming work to seek for the corresponding entanglement monotones.

\section{Acknowledgements}

This work was supported by the National Natural Science Foundation
of China, under Grant No. 10805007 and No. 10875020, and the
Doctoral Startup Foundation of Liaoning Province.

\end{document}